\documentclass[sigconf,nonacm,balance=false]{acmart}
\usepackage{smwifclass}
\usepackage[utf8]{inputenc}
\usepackage{amsmath}
\usepackage{amsfonts}
\unlessclass{llncs}{\usepackage{amsthm}}
\usepackage{algorithm}
\usepackage[endLComment=~]{algpseudocodex}

\usepackage{xcolor}
\usepackage{url}
\usepackage{smwmath}
\usepackage{smwtools}
\begin{IfCloudArxiv}
 \renewcommand{\spacetuneclass}[2]{#2}    
 \renewcommand{\bibliography}[1]{\input{#1.bbl}}  
 \newcommand{\TitleSpace}{\vspace{1.5cm}} 
 \newcommand{\PageOneBreak}{\pagebreak}
\end{IfCloudArxiv}
\begin{UnlessCloudArxiv}
  \newcommand{\TitleSpace}{\relax}        
  \newcommand{\PageOneBreak}{\relax}
\end{UnlessCloudArxiv}
\DeclareMathOperator{\iprec}{prec}
\DeclareMathOperator{\ishinv}{shinv}
\DeclareMathOperator{\ishift}{shift}
\DeclareMathOperator{\ifrac}{frac}
\DeclareMathOperator{\irev}{rev}
\DeclareMathOperator{\idegree}{degree}
\newcommand{\NNInt}{\ensuremath{\mathbb N}}

\newcommand{\zival}[4]{\ensuremath{#1#3\,..\,#4#2}} 
\newcommand{\SZ}{\ensuremath{S_{\mathbb Z}}}
\newcommand{\SR}{\ensuremath{S_{\mathbb R}}}

\newcommand{\iter}[1]{\ensuremath{_{(#1)}}}

\newcommand{\Rofx}{\ensuremath{R[x]}}
\newcommand{\Fofx}{\ensuremath{F[x]}}
\newcommand{\ldig}[1]{\ensuremath{_{[#1]}}}
\newcommand{\Sc}[1]{\ensuremath{\text{\sc #1}}}
\newcommand{\captionstrut}{\upstrut{2.0ex}}
\newcommand{\algosmallspace}{\upstrut{2.8ex}}
\newcommand{\algomedspace}{\upstrut{3.5ex}}

\newcommand{\algokw}[1]{{\bf #1}}
\newcommand{\Uses}[1]{\newline\indent{\hspace{-1.28cm}}\algokw{Uses:}\hspace{3.8mm} #1}

\begin{IfClassAcm}
   \copyrightyear{2023}
   \acmYear{2023}
   \setcopyright{acmlicensed}
   \acmConference[ISSAC 2023]{International Symposium on Symbolic and Algebraic Computation 2023}{July 24--27, 2023}{Troms\o, Norway}
   \ifx\acmBooktitle\undefined\def\acmBooktitle #1\relax\fi
   \acmBooktitle{International Symposium on Symbolic and Algebraic Computation 2023 (ISSAC 2023)}
   \acmPrice{15.00}
   \acmDOI{10.1145/3597066.3597076}
   \acmISBN{979-8-4007-0039-2/23/07}
   \orcid{0000-0001-8303-4983}
\end{IfClassAcm}
\begin{document}
\title{Efficient Generic Quotients Using Exact Arithmetic}
\author{Stephen M. Watt}
\newcommand{\theabstract}{
\begin{abstract}
The usual formulation of efficient division uses Newton iteration to compute an inverse in a related domain where multiplicative inverses exist. 
On one hand, Newton iteration  allows quotients to be calculated
using an efficient multiplication method. On the other hand, working in another domain is not always desirable and can lead to a library structure where arithmetic domains are interdependent.
This paper uses the concept of a whole shifted inverse and modified Newton iteration to compute quotients efficiently without leaving the original domain. 
The iteration is generic to domains having a suitable shift operation, such as integers or polynomials with coefficients that do not necessarily commute.
\end{abstract}
}
\begin{IfClassAcm}
  \affiliation{
    \institution{Cheriton School of Computer Science,
                 University of Waterloo}
    \city{Waterloo}
    \country{Canada\ifCloudArxiv\TitleSpace\fi}
  }
\end{IfClassAcm}
\begin{IfClassLlncs}
  \institute{Cheriton School of Computer Science, University of Waterloo\\
  \url{http://www.springer.com/gp/computer-science/lncs} \\
  \email{smwatt@uwaterloo.ca}}
\end{IfClassLlncs}
\begin{IfClassArticle}
  \date{
    Cheriton School of Computer Science\\
    University of Waterloo\\
    \texttt{smwatt@uwaterloo.ca}
  }
\end{IfClassArticle}
\EarlyTitleAbstract
\begin{IfClassAcm}
  \ccsdesc[500]{Theory of computation~Design and analysis of algorithms}
  \ccsdesc[500]{Mathematics of computing~Mathematical software}
  \ccsdesc[500]{Computing methodologies~Symbolic and algebraic algorithms}
  \ccsdesc[500]{Computing methodologies~Computer algebra systems}
  \keywords{quotient, remainder, integer arithmetic, polynomial arithmetic, modified Newton iteration, generic algorithms, library structure}
\end{IfClassAcm}
\LateTitleAbstract

\section{Introduction}
\label{sec:introduction}
Multiple precision integer arithmetic and polynomial arithmetic lie at the heart of a number of computational fields, including computer algebra and cryptography.
The most fundamental operations that cannot generally be performed in time linear in the size of the inputs are multiplication and division, \textit{i.e.} quotient or remainder.  Efficient algorithms for these operations are therefore important.

One method to perform fast integer division  is to compute the inverse of the divisor to sufficient precision by Newton iteration on approximate real numbers and then obtain the quotient by multiplication.  The products in the iteration step and the final one  can be performed by fast multiplication to give fast division.  
This approach requires working in some model of the real numbers such as multiple precision floating point arithmetic, which may be undesirable.
Fast computation of univariate polynomial quotients may be performed using an ideal-adic Newton iteration on reverse polynomials in $F[x]/\langle x^n\rangle$. This allows the definition of an inverse and an algebraic mechanism to drop what would be low-order terms in a direct formulation.
In both the integer and polynomial cases, these methods leave the original domain.  This can complicate library structure and obscure potential optimizations.

\spacetune{\PageOneBreak}
Consider the consequences of using an approximation to the divisor inverse in computing integer quotients.
Basic integer operations now require a representation for approximate real numbers, either as 
multiple precision floating point or by some implicit  mechanism. 
This encourages a structure where approximate and exact arithmetic are mutually dependent.
Of course, extended precision floating point arithmetic libraries ultimately use integer operations, so it could be argued that they are integer computations, but the approach is significantly different.
Algorithms in models of real arithmetic typically rely on values being smaller than small relative error bounds.
In floating point arithmetic this is often phrased in terms of number of units in the last place, or ``ulps''.
This is quite different than the exactness required for integer arithmetic, and totally ignores the arithmetic dynamics questions of  integer iterations.

\vfill
This paper presents an alternative direct iteration 
that can be 
formulated generically on rings with an efficient shift operation. 
Arithmetic is exact, remaining in the original ring and without increasing computational complexity.
We show how an iterative method may be used to compute a ``whole shifted inverse''.
This quantity can then be used to compute the quotient and remainder.  
The algorithm relies on multiplication, the method for which can be given as a parameter.   Thus, even when fast multiplication relies on other abstractions, the core arithmetic library will not have a dependency.

\vfill
The contributions of this paper are:
\spacetuneclass{acmart}{\vspace{-0.2\baselineskip}}
\begin{itemize}
    \item generic whole shift and shifted inverse as basic operations,
    \item an in-domain iterative method for the whole shifted inverse, 
    \item an analysis of the integer iteration properties,  proofs of convergence, useful bounds and quantitative results,
    \item an efficient generic algorithm for quotients.
\end{itemize}

\vfill
The remainder of this paper is organized as follows:
We begin in Section~\ref{sec:background}
with  some notational conventions, basic facts and relevant complexity results.
Section~\ref{sec:whole-shifted-inverse} presents
the concept of the whole shifted inverse for both integers and polynomials and introduces iterative methods based on it.
Section~\ref{sec:integer-iteration-properties} analyzes the behaviour of the integer iteration, and in particular shows where it has fixed points and the number of steps to arrive at one.  
Section~\ref{sec:base-and-precision-matters} shows how these results can be applied to selecting the starting point and limit the size of intermediate computations.
Section~\ref{sec:integer-algorithm} combines the results of the analysis to give a complete integer algorithm. Section~\ref{sec:generic-algorithm} presents the algorithm in a generic form and applied to polynomials. 
Finally, Section~\ref{sec:conclusions} provides some concluding remarks.
\spacetuneclass{acmart}{\pagebreak}

\spacetuneclass{acmart}{\vspace{-0.2\baselineskip}}
\section{Background}
\label{sec:background}
\subsubsection*{\textbf{Notation}}
In addition to conventional notation, we adopt the following:
\begin{center}
\begin{tabular}{@{}cl@{}}
$\zival()ab$,
$\zival[)ab$,
\textit{etc}
    & real intervals intersected with $\mathbb Z$ \\
$ u \iquo v,\; u \irem v$ 
    & quotient and remainder (see below)  \\
$\iprec_B u$
    & number of base-$B$ digits, $\floor{\log_B |u|} + 1$\\
$\iprec_x p$
    & number of coefficients, $\idegree_x p + 1$\\
$ \ifrac x$
    & fractional part, $x - \floor x$ \\
$\ishift_{n,X} v$
    & whole shift  (see Section~\ref{sec:whole-shifted-inverse})\\
$\ishinv_{n,X} v$
    & whole shifted inverse (see Section~\ref{sec:whole-shifted-inverse})\\
$X\iter i$  
    & value of $X$ at $i^{th}$ iteration
\end{tabular}
\end{center}
The integer interval notation, ``[a..b]'' \textit{etc}, is used by Knuth, \textit{e.g.}~\cite{Knuth-vol4b}.
The ``$\iprec$'' notation, abbreviating ``precision'', is similar to that of~\cite{Moenck1972} where it is used to present certain algorithms generically for integers and polynomials.
We take integers to be represented in base-$B$.  That is, for any integer $u \ne 0$ there is $h = \iprec_B(u)-1$, such that
\(
u = \sum_{i=0}^h u_i B^i, \quad u_i \in \mathbb Z,\, 0 \le u_i < B, \; u_h \ne 0.\)

\subsubsection*{\textbf{Division}}
Given $u, v \in D$ for $D$ an integral domain with Euclidean norm \mbox{$N\!\!: D \rightarrow \NNInt$}, there exist quotient $q$ and remainder $r$ in $D$ such that
\( 
u = q\,v + r,\; r = 0~\text{or}~N(r) < N(v).
\)
For $D$ being $\mathbb Z$ with $N = \text{abs}$
or $F[x]$, $F$ a field, with $N = \idegree$, the quotient and remainder are unique and we write $q = u \iquo v$ and $r = u \irem v$.

Given $u, v \in \mathbb R$, $v > 0$, we often use the whole quotient $q$ and fractional remainder $r$, these being $q = \floor{u/v}$ and $r = u - q\,v$.   
For~$u, v \!\in\! \mathbb R$, $x\iter 0\! \in\! (0, 2u/v)$, the value $u/v$ for easily invertible $u$ is the solution to $f(x) = u/x - v = 0$, computed by Newton iteration
\begin{equation}
x\iter{i+1} = x\iter i - \frac{f(x\iter i)}{f^\prime(x\iter i)} = x\iter i + x\iter i \left (1 -\frac vu x\iter i \right ),\quad x\iter i \in \mathbb R.
\label{eqn:newton-division-iteration}
\end{equation}

\subsubsection*{\textbf{Algorithms}}
When efficient division uses multiplication the computational complexities of the two operations are intimately related.
The classical algorithms for multiplication
and division of $N$-bit integers require time $O(N^2)$~\cite{knuth-vol2}.
The best known upper bound for multiplication complexity is $O(N \log N)$~\cite{H+vdH-2021} and this is believed to be tight~\cite{afshani+-2019}.
While this gives the best asymptotic behaviour, it is not not suitable for practical use.
In practice, software libraries such as GMP~\cite{gmp} use different methods for different size inputs.
For multiplication, typically the $O(N^2)$ classical method is used for smallest values, the Karatsuba $O(N^{\log_2 3})$ method~\cite{karatsuba} or a Toom-Cook $O(N^{\log_3 5})$ method~\cite{cook-phd} for intermediate sized values, and another method, such as the Sch\"onhage-Strassen $O(N \log N \log \log N)$ FFT method~\cite{SS-FFT}, for the largest values.

Using a multiplication with complexity $O(M(N))$, polynomial division may be computed by Newton iteration in complexity at most $O(\log N \,M(N))$ or complexity $O(M(N))$ if fast multiplication is used.~\cite{bernstein-fastmult}.
For integer division, working with extended precision floating point, matters are slightly more complicated due to carries and the region of convergence.
Aho, Hopcroft and Ullman~\cite{aho-hopcroft-ullman} show how an approach similar to ours may be used for integers base 2.
Hitz and Kaltofen~\cite{HitzKaltofen} show how Newton iteration may be used to compute reciprocals in residue number systems.

It is often preferable to use
classical $O(N^2)$ division methods for small values and direct methods with Karatsuba complexity, such as that of Jebelean~\cite{jebelean-div2} or Burnikel and Ziegler~\cite{Burnikel+-1998}, for integers of intermediate size.  Useful overviews are given in~\cite{bernstein-fastmult,GGR2020}.
\spacetuneclass{acmart}{\pagebreak}
\section{The Whole Shifted Inverse}
\label{sec:whole-shifted-inverse}
\subsection{Integer Definitions and Facts}
We are interested in computing quotients and remainders of integers  using ring operations, rather than in a model of the reals.
We make use of an operation that is either a multiplication or a special weak quotient, that is:
\begin{Definition}[Whole shift in $\mathbb Z$]
Given integers $B > 1$, $n$ and $u$, the \emph{base-$B$ whole $n$-shift of $u$} is
\begin{equation*}
    \ishift_{n,\, B}(u) = \floor{u B^n}.
\end{equation*}
When $B$ is clear by context, we write $\ishift_n u$.
\end{Definition}
When $n \ge 0$, this the integer multiplication $u \times B^n$. When $n < 0$, this is a specialized quotient.   
If integers are given in base-$B$, this operation can be highly efficient, taking time $O(1)$ or $O(\log u+n)$, depending on the details of the representation.  

We now define another specialized quotient, the computation of which is the main object of this article. 
\begin{Definition}[Whole shifted inverse in $\mathbb Z$]
Given integers $B > 1$, $n\ge 0$ and $v\ne 0$, the \emph{whole base-$B$ $n$-shifted inverse of $v$ with respect to $B$} is
\begin{equation*}
    \ishinv_{n,\,B} (v) = \floor{B^n/v}.
\end{equation*}
When $B$ is clear by context, we write $\ishinv_n v$.
\end{Definition}
This operation  generalizes the {\sc reciprocal} operation of~\cite{aho-hopcroft-ullman} to arbitrary bases,
$
    \ishinv_{2\log_2 v-1,\,2}(v) = \text{\sc reciprocal}(v),
$
and it can be used to compute general quotients
in what can be viewed as a case of Barrett reduction~\cite{barrett-1986,hasenplaugh-2007}.

\begin{Theorem}[Quotient by whole shifted inverse in $\mathbb Z$]
    Given two positive integers $u$ and $v$,
    with $u \le B^h$, 
\begin{equation*}
    u \iquo v = \ishift_{-h}(u \cdot \ishinv_h v) + \delta, \quad \delta \in \{0, 1\}.
\end{equation*}
\end{Theorem}
\begin{Proof}
From the definitions and the fact that $u \le B^h$, we have
\begin{align*}
    \ishift_{-h}(u \cdot \ishinv_h v) 
    &=  
    u/B^h (B^h/v -\epsilon_1 ) - \epsilon_2,
    \quad 0 \le \epsilon_i < 1 
\end{align*}
\begin{equation*}
    \Leftrightarrow
    u \iquo v -2
    < 
    \ishift_{-h}(u \cdot \ishinv_h v) 
    \le
    u \iquo v + (u \irem v)/v.
\end{equation*}
Since $(u\irem v)/v < 1$ and  $\ishift$ maps to $\mathbb Z$, the result follows.
\end{Proof}
Checking all $2 \le u \le 10^6$, $2 \le v \le u$, we find that $\delta = 0$ and $\delta = 1$ occur with approximately equal frequency. 

\subsection*{An Integer Iteration to Compute $\ishinv_h v$}
We observe that if $u$ is specialized to $B^h$ in the Newton iteration \eqref{eqn:newton-division-iteration}, then it becomes an iteration to compute $\ishinv_h v$.  
This iteration requires a real division, however.   This real division is close to being a shift, so we instead use the modified iteration:
\begin{align}
    w\iter{i+1} &= w\iter i + \ishift_{-h} \big (\ishift_h w\iter i - v w\iter i^2 \big),
            \quad\quad w\iter i \in \mathbb Z
            \label{eqn:integer-iteration}
            \\
            &= w\iter i + \Floor{w\iter i (B^h - v w\iter i) B^{-h}}
            \nonumber
\end{align}
and show in Section~\ref{sec:integer-iteration-properties} that the iteration gives the desired result
\begin{equation*}
    w\iter i \rightarrow \ishinv_h v.
\end{equation*}
We note that this is \emph{not} the usual Newton iteration, as it discretized to integers.  We must therefore examine its properties in order to justify our claim that it computes the whole shifted inverse.

\subsection{Polynomial Definitions and Facts}
We are also interested in the efficient computation of univariate polynomial quotients.   A well-known method is to use Newton iteration to compute a modular inverse of a reversed polynomial.   Specifically, to compute $q = u \iquo v$ for $u, v \in F[x]$, let  $h$ and $k$ be the degrees of $u$ and $v$ respectively, and
\begin{align*}
v^* &= \text{inverse of }\irev_k v &&\pmod{x^{h-k+1}} \text{  by Newton iteration} \\
q^* &= \irev_h u \times v^*        &&\pmod{x^{h-k+1}} \\
q     &= \irev_{h-k} q^*
\end{align*}
where
\(
\irev_n p(x) = x^n p(1/x).
\)
This is detailed nicely in~\cite{vzgg-mca}.    The use of reverse polynomials modulo $x^{h-k+1}$ is used to drop low-order terms.   This reversal trick does not work for integer quotients because carries would propagate in the wrong direction.   
This was the reason to formulate the integer iteration in terms of $\ishift$ and $\ishinv$.
These operators may be defined analogously for polynomials to give an iteration without reversals.  This direct formulation has the benefit that it admits  certain optimizations, as shown in Section~\ref{sec:generic-algorithm}.
\begin{Definition}[Whole shift in \Rofx]
Given a polynomial $u = \sum_{i=0}^h u_ix^i \in R[x]$ and integer $n$,  the \emph{variable-$x$ whole $n$-shift of $u$} is
\begin{equation*}
\ishift_{n,x} u = \sum_{i+n\ge 0} u_i x^{i+n}.
\end{equation*}
When $x$ is clear by context, we write $\ishift_n u$.
\end{Definition}
\begin{Definition}[Whole shifted inverse in $\Fofx$]
Given $n \in \NNInt$ and $v\in \Fofx$ with field $F$, the \emph{whole  $n$-shifted inverse of $v$ with respect to $x$} is
\begin{equation*}
   \ishinv_{n,x} v = x^n \iquo v.
\end{equation*}
When $x$ is clear by context, we write $\ishinv_n v$,
\end{Definition}
With these definitions,we have the following simple theorem.
\begin{Theorem}[Quotient by whole shifted inverse in $\Fofx$]
Given two polynomials $u, v \in \Fofx$ with field $F$  and $0 \le \idegree u \le h$,
\begin{equation*}
u \iquo v = \ishift_{-h}(u \cdot \ishinv_h v).
\end{equation*}
\end{Theorem}
\begin{Proof}
Letting $p[i]$ denote the coefficient of $x^i$ in $p$,  we have, for $i \in \zival[]0{h-\idegree v}$,
\begin{align*}
 (u \iquo v)[i] &= \big ((x^hu) \iquo v\big)[i+h] =\big (x^{-h}  u  (x^h \iquo v)\big)[i] \\
                       &= \big ( \ishift_{-h}(u \cdot \ishinv_h v) \big)[i],
\end{align*}
giving the desired result.
\end{Proof}
\subsection*{A Polynomial Iteration to Compute $\ishinv_h v$}
The polynomial iteration to compute $y = \ishinv_h v$ corresponding to~\eqref{eqn:integer-iteration} takes the same form, 
\begin{equation}
 y\iter{i+1} = y\iter i + \ishift_{-h} \big (\ishift_h y\iter i - v {y\iter i}^2 \big),\quad y\iter i \in F[x].
\label{eqn:polynomial-iteration}
\end{equation}
This is a direct formulation of the method with reverse polynomials.
\section{Integer Iteration Properties}
\label{sec:integer-iteration-properties}
The convergence of Newton iteration on the reals is well understood.  We are interested, however, in the iteration of an integer-valued function involving fractions, subtraction and rounding, so some care is required to ensure that the arithmetic dynamics do not give unexpected phenomena.
We consider the two functions
\begin{align}
    \SR: \mathbb R \rightarrow \mathbb R &= x \mapsto x + x \left (1- \frac vu x \right ) \\
    \SZ: \mathbb Z \rightarrow \mathbb Z &= w \mapsto w + \Floor{ w \left ( 1 - \frac vu w \right ) }, \quad 1 < v < u.
    \label{eqn:sz}
\end{align}
where it is sometimes more convenient to use these in the form
\begin{align}
    \SR(x) &= x\left (2 - \frac vu x \right ) \\
    \SZ(w) &= \Floor{\SR(w)}.
\end{align}
The iteration $w\iter{i+1} = \SZ(w\iter i)$ gives \eqref{eqn:integer-iteration}  when $u = B^h$.   We do not specialize $u$ in the present analysis, however, as the properties of the integer iteration do not depend on $u$ having any particular form.
\subsection{Real Convergence}
We first show the properties of $\SR$ to provide an orientation for the study of $\SZ$.  
We begin with the following simple real theorem for which the integer case is less straightforward.
\begin{Theorem}[$\SR$ Fixed Points]
The function $\SR$ has fixed points $0$ and $u/v$ and no others.
\end{Theorem}
\begin{Proof}
If $\SR(x) = x$, then
\begin{equation*}
                x + x \left ( 1 - \frac vu x \right )   = x 
    \Leftrightarrow x \left ( 1 - \frac vu x \right )   = 0
\end{equation*}
and the result follows.
\end{Proof}
\noindent
We will make use of the following result.
\begin{Theorem}[$\SR$ Iterates]
\label{thm:sr-iterates}
The iterates \/$\SR^i$ of \/ $\SR$ are given by 
\begin{equation}
\SR^i(x) = \frac uv \left ( 1- \left ( 1-  \frac vu x\right )^{2^i} \right ), \quad i \ge 0.
\label{eqn:SRiterformula}
\end{equation}
\end{Theorem}
\begin{Proof} 
We use induction on $i$.  When $i = 0$, equation \eqref{eqn:SRiterformula} is satisfied:
\begin{equation*}
    \SR^0(x) = x = \frac uv \left (1 - \left (1 - \frac vu x\right )^{2^0} \right ).
\end{equation*}
If for some $i = n \ge 0$ equation \eqref{eqn:SRiterformula} holds, then 
\begin{align*}
 \SR^{n+1}(x) &= \SR^n(x) \left ( 2 - \frac vu \SR^n(x) \right )  \\
 &= \frac uv \left ( 1- \left(1-\frac vu x\right )^{2^n} \right ) \cdot \left ( 1 +\left(1+\frac vu x\right )^{2^n} \right ) \\
 &= \frac uv \left ( 1- \left(1-\frac vu x\right )^{2^{n+1}} \right )
\end{align*}
and equation \eqref{eqn:SRiterformula} also holds for $i = n+1$.
\end{Proof}

\noindent 
The following theorem describes how iterates behave at all points on the real line.
\begin{Theorem}[$\SR^i$ Convergence]
\label{thm:realconvergence}
The sequence of iterates \;$\SR^i(x), i \ge 1$   converges if and only if $x \in [0, 2u/v]$.  If $x = 0$ or $x = 2u/v$, then $\SR^i(x) = 0, i \ge 1$. If $x \in (0, 2u/v)$, the sequence converges quadratically to $u/v$.
\end{Theorem}
\noindent
\begin{Proof}
The proof is split into disjoint and exhaustive cases:

\Case{Case 1, $x = 0$ or $x = 2u/v$} We have $\SR(x) = 0$ so $\SR^i(x) = 0, i \ge 1$.

\Case{Case 2, $x < 0$} When $uv \ge 0$,  we have $-x^2v/u \le 0$ so $\SR(x) = 2x -x^2u/v \le 2x < 0$.   When $uv < 0$, we have $2- xv/u > 2$ so $\SR(x) = x(2-xv/u) < 2x$.  In either case, $\SR^i(x) \le 2^i x$ and grows negatively without bound.

\Case{Case 3, $x > 0, uv \le 0$}  Since $2x > 0$ and $-x^2 v/u  \ge 0$, we have \( \SR(x) = 2x -x^2 v/u \ge 2x\) and \( \SR^i(x) \ge 2^i x\) and grows without bound.

\Case{Case 4, $0 < x < 2u/v, uv > 0$}
We observe that $ 0 < \SR(x) \le u/v$ in this region.   To see this, note $\SR(x)$ is a parabola with maximum $\SR(x) = u/v$ at the vertex $x=u/v$, and value 0 at the excluded region endpoints.   So $\SR$ maps the entire region into $(0, u/v]$ and we need only consider $\SR^i\big ( (1-\epsilon) u/v \big )$ with $0 \le \epsilon < 1$.  By Theorem~\ref{thm:sr-iterates},
\begin{equation*}
\SR^i \left ((1-\epsilon)\frac uv\right) = (1-\epsilon)^{2^i} \frac uv, \quad i \ge 0
\end{equation*}
so $\SR^i$ converges in this region and, moreover,
\begin{align*}
    \frac{\left | \SR^{i+1}(x)- u/v \right|}{\phantom{^\lambda}\left | \SR^i(x)-u/v\right |^\lambda} 
    \detail{
        = \frac{\left |  
                 \left ( 1- \frac vu x\right )^{2^{i+1}} 
        \right|}{\phantom{^2}\left| 
                 \left (1- \frac vu x\right )^{2^{i}}
        \right |^2 }} = v/u \quad \text{for }\lambda = 2.
\end{align*}
Therefore $\SR^i(x)$ converges quadratically to $u/v$ for $ 0 < x < 2u/v$.

\Case{Case 5, $ x > 2u/v,  uv > 0$}
We have $2 - xv/u < 0$ so 
\(
\SR(x) = x\left ( 2 -  \frac vu x\right) <  0
\)
and $\SR^i(\SR(x))$ grows negatively without bound by case 2.

\Case{Summary} Cases 1 and 4 show that the sequence converges to the claimed values when $x \in [0, 2u/v]$. Cases 2, 3 and 5 together show that the sequence does not converge when $x \not \in [0, 2u/v]$.  Note $[0, 2u/v]$ is empty when $uv < 0$.
\end{Proof}

\subsection{Integer Fixed Points}

\begin{figure}[t]
    \centering
    \includegraphics[width=0.905\columnwidth]{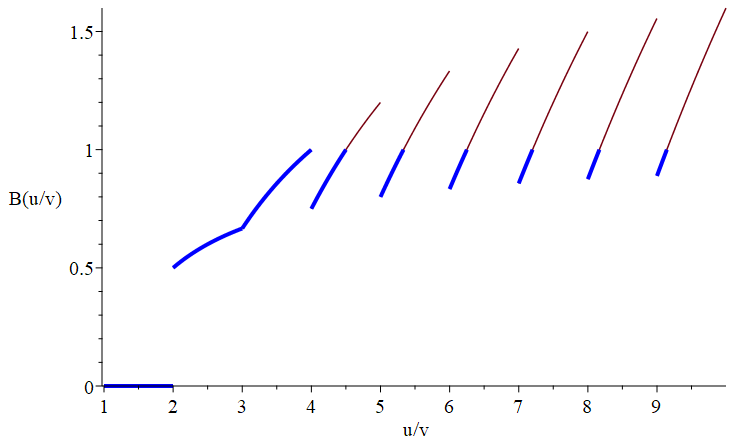}
\spacetuneclass{acmart}{\vspace{-11
pt}}
    \caption{Region where $\SZ(\floor{u/v}-1) = \floor{u/v} -1$ (heavy line).}
    \label{fig:Buv}
\end{figure}

In order to study convergence of the sequence of $\SZ$ iterates, 
we first show where $\SZ$ has fixed points.
\begin{Theorem}[$\SZ$ Fixed Points]
\label{thm:integer-fixedpts}
Given $1 < v < u \in \mathbb Z$,
the fixed points of $\SZ$  on $\zival[]0{u/v}$
are 0, 1, $\floor{u/v}$ and, when 
\begin{equation}
  \frac uv \in (1,4) \cup  \bigcup_{j=4}^{\floor{u/2}} \left [ \, j, \; j + \frac 1{j-2} \right ),
  \label{eqn:uvm1-condition}
\end{equation}
$\floor{u/v}-1$.  These are 2, 3, or 4 distinct points, depending on the value of $u/v$.
\end{Theorem}
\begin{Proof}
The values 0, 1 and $\floor{ u/v }$ are easily seen to be fixed points:
\begin{align*}
    \SZ(0) &= 0 + \floor 0  = 0 \\
    \SZ(1) &= 1 + \floor{1 - v/u} = 1\\
    \SZ(\floor{u/v}) 
           &= \floor{u/v} + \floor{ \ifrac(u/v) \big(1 - \ifrac(u/v)\, v/u\big) } = \floor{u/v} \\
           &\phantom = \text{since } u > v.
\end{align*}
We find \(
     \SZ(\floor{u/v}-1) = \floor{u/v}-1  \)
is equivalent to 
\begin{equation}
     0 \le B(u/v) < 1, \text{ where } B(x) = (\floor{x}-1) (x - \floor{x} + 1)/x.
     \label{eqn:uvm1-precondition}
\end{equation}
This is satisfied for $1 < u/v < 4$ because $B(x) \ge 0$ and $B(u/v) < 1$ is equivalent to $v/u = \ifrac (u/v)/\big( 1+\ifrac (u/v)\big )^2$. For $u/v \ge 4$, we have $B(u/v) - 1 = j-2 - (j-1)^2 v/u$ on $[j, j+1), j \in \mathbb Z$. Therefore $B(u/v) < 1$ on $[j, j+1)$ when $u/v \in \big [j, j+1/(j-2) \big )$.
Since $v \ge 2$, condition~\eqref{eqn:uvm1-precondition} holds exactly when
\eqref{eqn:uvm1-condition} is satisfied,
so $\floor{u/v}-1$ is a fixed point of $\SZ$ if and only if \eqref{eqn:uvm1-condition} is satisfied.

We now show there are no other integer fixed points.  
Fixed points must satisfy
\begin{equation}
\SZ(w) - w = 0 \Leftrightarrow
    \Floor{w \left ( 1 - \frac vu w \right)} = 0 
    \Leftrightarrow
    0 \le w - \frac vu w^2 < 1.
    \label{eqn:fixed-pt-condition}
\end{equation}
This locus  lies below the parabola $E(x)=x-(v/u) \, x^2$ which is symmetric about $x= u/(2v)$, \textit{i.e.} $E(u/(2v) - a) = E(u/(2v)+a)$, with $E$ increasing for $x < u/(2v)$ and decreasing for  $x > u/(2v)$. 

We separately consider the points to the left and right of the line of symmetry.
Only when $u \ge 4v$ is $\zival[]0{u/v} - \{0, 1, \floor{u/v}-1, \floor{u/v}\}$ non-empty.  
Consider the elements $2 \le n \le \floor{u/(2v)}$. 
Since $E$ is increasing on this region and $u \ge 4v$,
\begin{equation*}
    E(w) \ge E(2) = 2- \frac vu 4 \ge 1,
\end{equation*}
showing these values do not satisfy \eqref{eqn:fixed-pt-condition} so are not fixed points. Next consider the elements $w \in  D$ with $\floor{u/(2v)} \le w \le \floor{u/v}-2$. Since $E$ is decreasing on this region, the smallest value of $E(w)$ will be $E(\floor{u/v} -2)$.   If $\floor{u/v}-2 \ne 2$, symmetry about $x = u/(2v)$ gives
\begin{align*}
   E(\floor{u/v}-2) 
         & = E(u/v - \floor{u/v} + 2)  \ge E(2) \ge 1,
\end{align*}
showing these values are also not fixed points.
Combining the two halves, $\SZ$ has no fixed points $n$ with $2 \le n \le \floor{u/v}-2$.

The cardinality of the set $\{0, 1, \floor{u/v}-1, \floor{u/v} \}$ will therefore be 2, 3 or 4, depend on the value of $u/v$ and the condition \eqref{eqn:uvm1-condition}.
\end{Proof}
The region where $0 \le B(u/v) < 1$ is shown in Figure~\ref{fig:Buv}.

\begin{Estimate}
\label{est:uvm1}
Given an integer $u > 2$, the number of integer values $v$, $1 < v < u$, for which $\SZ(\floor{u/v}-1) = \floor{u/v}-1$ is approximately
\begin{equation*}
     \frac{\pi^2-5}6 u.
\end{equation*}
\end{Estimate}
\begin{Proof}[Justification]
We estimate the frequency with which $\floor{u/v}-1$ is a fixed point of $\SZ$.  
The condition \eqref{eqn:uvm1-condition} is satisfied either when $u/v < 4$ or when $u/v$ lies in an interval $\big [i, 1/(i-2)\big )$ out of a total of $u-2$ values.  
We assume the values of $\ifrac(u/v)$ are uniformly distributed on $[j,j+1)$, and estimate the number of values of $v$ that fall in an interval to be proportionate to the length of that interval times the number of $u/v$ with $\floor{u/v} \in [j-1, j)$. This gives the estimate
\begin{align*}
{\Large \#}  \bigg \{\, 
   v \; \bigg |\; 
   & \SZ \left (\Floor{\frac uv}-1 \right) = \Floor{\frac uv}-1,\;  1 < v < u \bigg \}  \\
  &\approx u - \floor{u/4} - 1 + (u-2) \cdot \sum_{j = 4}^{\floor{u/2}} \frac1{j-1}\cdot \left (\frac 1{j-1} -\frac 1j \right )  \\
   &=  u - \floor{u/4} - 1 + (u-2) \cdot \left ( \Floor{\frac u2}^{-1}\!\!\!\! - \Psi \left (1,\Floor{\frac u2}\right ) - \frac{19}{12} + \frac{\pi^2}6\right) 
\end{align*}
and the result follows.
Here, $\Psi$ is the polygamma function, which makes a negligible contribution.
\end{Proof}
Figure \ref{fig:uvm1table} compares, for increasing $u$, the actual and estimated number of $ 1 < v < u$ making $\floor{u/v}-1$ a fixed point of $\SZ$.  


\subsection{Integer Convergence}
We now analyze the convergence of the integer iteration.  On sequences of integers, convergence means arrival at a fixed point rather than entering a cycle or growing unboundedly.

\begin{Theorem}[$\SZ$ Convergence]
\label{thm:integer-convergence}
The sequence of iterates \;$\SZ^i(w), i \ge 1$   converges if and only if $w \in \zival[]0{2u/v}$. If the series converges, then it is to one of $\{ 0, 1, \floor{u/v}-1, \floor{u/v} \}.$  For $w \in \zival[]2{u/v}$, $\SZ^i(w)$ converges to  $\floor{u/v}-1$ or $\floor{u/v}$.
\end{Theorem}

\begin{Proof}
The proof is organized in disjoint and exhaustive cases along the same lines as Theorem \ref{thm:realconvergence}.
Here, however, the cases are not independent. Their relationship is as follows:
case 1b depends on 1a, 4a and 4b,
case 4b depends on 4a, and the remaining
cases, 1a, 2, 3 and 4a, do not depend on others.   

\Case{Case 1a, $w = 0 \; \text{or}\;  1$}
If $w = 0 \text{ or } 1$, we have
\begin{equation*}
\SZ^i(w) =  w \in \{0, 1\}, i \ge 0.
\end{equation*}

\Case{Case 1b, $w = \floor{2u/v}-1 \; \text{or}\; \floor{ 2 u/v }$}
If $w = \floor{2u/v} = 2u/v - \delta_0$, then 
\begin{equation*}
 \SZ(\floor{ 2u/v }) = \Floor{  \delta_0 \left (2  -\delta_0 \frac vu \right ) }.
\end{equation*}
Since $\delta_0(2-2\delta_0v/u)$ is a concave-down parabola with vertex at $\delta_0 = u/v > 1$, it is increasing for $0 \le \delta_0 < 1$, taking values from $0$ to $2-v/u$.  Since $0 < v/u < 1$, we have
\begin{equation*}
    \SZ^i(\SZ(\floor{2u/v})) = \SZ(\floor{2u/v}) \in \{0, 1\}.
\end{equation*}
If $w = \floor{2u/v}-1 = 2u/v - \delta_1$, then
\begin{equation*}
    \SZ(\floor{2u/v} - 1) = \Floor{2\delta_1 + 2 - (\delta_1+1)^2 v/u}.
\end{equation*}
Again we have a concave-down parabola, now with vertex at $\delta_1 = u/v-1$ which may occur inside or outside $\in [0, 1)$, depending on $u/v$. Considering all cases, we have 
\begin{equation*}
    \SZ(\floor{2u/v}-1) \in \{0, 1, 2, 3\}.
\end{equation*}
The values 2 and 3 can arise only when $\floor{2u/v} \ge 4.$   For all $\floor{2u/v} \ge 4$, the sequences $\SZ^i(w)$ converge by cases 1a and 4a/b for $w \in \{0, 1, 2\}$.  If $\floor{2u/v} > 4$, then $\SZ^i(3)$ also converges by case 4a/b.  If $\floor{2u/v} = 4$, then $ 2v \le u < 5v/2$ so $\SZ(3) = 1$, giving case 1a. 
\begin{figure}[t]
\begin{center}
\begin{tabular}{crr@{\hspace{0.75em}}c@{}c}
  $\textbf{\itshape{u}}$
& \textbf{Actual} 
& \textbf{Estimate \ref{est:uvm1}}
& \textbf{Abs Err}
& \textbf{Rel Err}\\
\hline
$10^1$   &             8 &              8 &   0 & $0$                   \\
$10^2$   &            85 &             81 &   4 & $4.706\times 10^{-2}$ \\
$10^3$   &           818 &            811 &   7 & $8.557\times 10^{-3}$ \\
$10^4$   &         8,135 &          8,116 &  19 & $2.336\times 10^{-3}$ \\
$10^5$   &        81,178 &         81,160 &  18 & $2.217\times 10^{-4}$ \\
$10^6$   &       811,655 &        811,600 &  55 & $6.776\times 10^{-5}$ \\
$10^7$   &     8,116,081 &      8,116,007 &  74 & $9.118\times 10^{-6}$ \\
$10^8$   &    81,160,153 &     81,160,073 &  80 & $9.857\times 10^{-7}$ \\
$10^9$   &   811,600,878 &    811,600,733 & 145 & $1.787\times 10^{-7}$ \\
$10^{10}$& 8,116,007,538 &  8,116,007,335 & 203 & $2.501\times 10^{-8}$ 
\end{tabular}
\spacetuneclass{acmart}{\vspace{-12pt}}
\end{center}
\caption{Number of $v$ with $\SZ(\floor{u/v}-1)$ fixed point.}
\label{fig:uvm1table}
\end{figure}

\Case{Case 2, $w < 0$}  We write the iteration as 
\begin{equation*}
\SZ(w) = w + \Floor{ w - vw^2/u } \le 2w - vw^2/u  < 2w
\end{equation*}
so $\SZ(w) < 2^i w\, \text{for }i \ge 0$, which grows negatively without bound.

\Case{Case 3, $w > 0, uv \le 0$}  This does occur since $u > v > 0$.

\Case{Case 4a, $1 < w \le \floor{u/v}$, $uv > 0$}
On the region $(1, \floor{u/v}-1)$, we have $\SR$ strictly increasing and,
by Theorem \ref{thm:integer-fixedpts}, $\SZ$ has no fixed points. We then have $\SZ$ strictly increasing and
\begin{equation*}
    \SZ(w) \ge w+1 \quad \text{for } 1 < w < \floor{u/v}-1. 
\end{equation*}
Since $\SR(x)$ achieves its maximum value at $u/v$, we have $\SZ(w) \in (1, \floor{u/v}]$.
Since the iterates increase by at least 1, there exists $i < w$ such that 
\begin{equation*}
\SZ^i(w) \in \big (1, \floor{u/v} \big] \;\backslash \;\big (1, \floor{u/v}-1 \big) = \big [\floor{u/v}-1, \floor{u/v} \big ].
\end{equation*}
Therefore the sequence converges to $\floor{u/v}-1$ or $\floor{u/v}$.

\Case{Case 4b, $\floor{u/v} < w < \floor{2u/v} - 1$, $uv > 0$}
We assume $u/v \ge 2$, otherwise the region is empty.
Since $\SR$ is strictly decreasing on this region, $\SZ(w)$ will be minimal at the end point $\floor{2u/v}-2$.
By symmetry about $u/v$, this is
\begin{align*}
    \SZ(\floor{2u/v}-2)
         &= \floor{\SR(\floor{2u/v}-2)}
= \floor{\SR(2+2u/v - \floor{2u/v})} \\
         &= \Floor{ (\floor{2u/v} - 2) (2 + 2 v/u - \floor{2u/v} v/u)} \ge 2.
\end{align*}
Since $\SZ$ is bounded above by $\floor{u/v}$, we have $\SZ(w) \in [2, \floor{u/v}]$ for $\floor{u/v} < w < \floor{2u/v} -1$ and the $\SZ^i(w)$ converges by case 4a.

\Case{Case 5, $w > \floor{ 2u/v }$, $uv > 0$}
Let $w = \floor{ 2u/v } + N = 2u/v +N -\ifrac(u/v)$ where $1 \le N \in \mathbb Z$.
We write the iteration as 
\begin{align*}
\SZ(w)
         &= \Floor{ -(N-\ifrac(u/v)) \left (2 + v/u(N-\ifrac(u/v))\right )}
        < 0
\end{align*}
so the sequence grows negatively without bound by Case 2.
\end{Proof}
\subsection{Initial Value and Fast Convergence}
\label{sec:init-val-and-convergence}
Given $u$ and $v$, it is desirable to find $\floor{u/v}$ in as few iterations as possible. 
To do so requires a good choice of starting value.  
If the starting value is too large, that is if it is larger than $\floor{2u/v}$, then the sequence of iterates will diverge.    If it is positive, but too small, then there is another problem.   Suppose that the result $\floor{u/v}$ has $b$ bits and the starting value has $b_0 < b$ bits.   Then there will be $b-b_0$ iterations to reach an iterate with the correct length since each iteration can multiply its argument by no more than 2. That is, there will be one iteration per bit short, while we expect the usual Newton iteration to double the number of correct digits.   This may be remedied with the following theorem.

\spacetuneclass{acmart}{\vspace{-0.25\baselineskip}}
\begin{Theorem}[Fast $\SZ$ Convergence]
\label{thm:fast-converge}
If $w\iter0 \in \zival{\big[}{\big]}{(1-\frac14)\, u/v}{(1+\frac14)\, u/v}$,
$u/v\ge 2$, then 
\begin{equation*}
    \SZ^{\ceil{\log_2 \log_2(u/v)}}(w\iter0) \in \big \{ \Floor{u/v} -1, \Floor{u/v} \big \}.
\end{equation*}
\end{Theorem}
\begin{Proof}
We begin by showing the claim holds for $\SR$.   
Since $u > v$ there will be at least one iteration, after which all values will be $x\iter i = \SR^i(x) \le u/v$.    
So we need only consider $(1-\frac14) u/v \le x \le u/v$.   
Then, for $\alpha \in \mathbb R$,
\begin{equation*}
    x\iter i = \left ( 1 - \frac 1{2^{2^\alpha}} \right) \frac uv
    \Rightarrow x\iter{i+1} = \SR(x\iter i) = 
     \left ( 1 - \frac 1{2^{2^{\alpha+1}}} \right) \frac uv.
\end{equation*}
When $x\iter0$ has the same number of bits as $\floor{u/v}$, the number of correct leading bits of $x\iter i$ doubles with each iteration.

We now consider $\SZ$.  At each iteration we have
\begin{equation*}
w\iter{i+1} = \SZ^{i+1}(w\iter0) = \SR(\SZ^i(w\iter0)) + \epsilon, \quad 0 \le \epsilon < 1.
\end{equation*}
If $w\iter i$ has $\alpha\iter i$ correct leading bits, then $w\iter{i+1}$ will have at least $2\alpha\iter i-1$ correct leading bits, so $\alpha\iter i = (\alpha\iter0 - 1)2^i+1$, with $\alpha\iter0 $ the number of correct leading bits of $w\iter0$.
Since $w\iter0 \in [(1-\frac14)u/v, (1+\frac14])u/v$, 
\begin{equation*}
\alpha\iter0 \ge \log_2(4-1) = \log_2 3 > 2.
\end{equation*}
 All bits will be correct if the number of iterations, $i$, satisfies
\begin{equation}
    (\alpha\iter0 -1) 2^i + 1 \ge \log_2(u/v)
    \label{eqn:alpha-condition}
\end{equation}
so when $i \ge  \ceil{\log_2 \log_2(u/v)}$, we have
$ 2^i +1 >  \log_2(u/v) $
and condition \eqref{eqn:alpha-condition} is satisfied. This gives the desired result.
\end{Proof}

\noindent
The condition $u/v \ge 2$ is given to avoid degenerate cases. Finding $\floor{u/v} = 1$ is easily achieved by testing $u < v+v$.

\section{Integer Base and Precision Matters}
\label{sec:base-and-precision-matters}
We now apply the previous results to computing $\ishinv_h v$ in base-$B$.  There are three questions to settle.  The first is how to obtain an iteration starting point efficiently that satisfies the conditions of Theorem~\ref{thm:fast-converge}.   The second is how to exploit iteration accuracy to perform intermediate computations on smaller quantities.  The third is to understand when a prefix of $v$ is sufficient to compute an iterate.  This section answers these questions.

\spacetuneclass{acmart}{\vspace{-3pt}}
\subsection{Initial Value}
We show a choice for the initial value of the iteration sequence that satisfy the
conditions of Theorem~\ref{thm:fast-converge}.
This involves inverting a short prefix of $v$.
As the necessary prefix size is bounded, this short inversion is a constant time operation.
\begin{Theorem}[Initial Value Choice]
Let  $B \le v < B^{k+1}$ and $2v \le u=B^h$ for $B \ge 16$ and
$v = V B^{k-f} + R$ with
$f, V, R \in \mathbb Z$,
$B^f \le V \!< B^{f+1}$, $0 \le R\! < B^{k-f}$
and $f \ge \min(k,2)$.
Then the choice  $w\iter0 = \floor{B^{f+2}/V} B^{h-k-2}$ gives
\begin{equation*}
\SZ^{\ceil{\log_2\log_2(u/v)}} (w\iter0) \in \{ \floor{u/v}, \floor{u/v} - 1 \}.
\end{equation*}
\end{Theorem}
\begin{Proof}
Since $V \ge 4$ and $R < B^{k-f}$, we have $\frac14 > R/(VB^{k-f})$ so
\[
  \left (1+\frac14 \right ) \frac uv
  > \left (1+\frac R{VB^{k-f}} \right ) \frac uv
  = \frac {B^h}{VB^{k-f}}
  \ge \Floor{\frac {B^{f+2}}V}B^{h-k-2}.
\]
On the other hand, since $V < B^{f+1}$ we have $B^{f+2}/(4V) > 1$ and
\begin{align*}
\left(1-\frac 14 \right) \frac uv 
&= \frac 34 \frac u{VB^{k-f}+R}
\le \frac 34 \frac {B^{h-k+f}}V  \\
&<  \left ( \Floor{\frac{B^{f+2}}V} + 1 - \frac{B^{f+2}}{4V}\right)
B^{h-k-2}
< \Floor{\frac {B^{f+2}}V} B^{h-k-2}.
\end{align*}
The conditions of Theorem~\ref{thm:fast-converge} are satisfied so we have our result.
\end{Proof}
If $B < 16$, one may interpret the value $v$ as base-$B^p$, which need not involve 
copying or modifying any data.

\subsection{Shorter Iterates}
\label{sec:shorter-iterates}

Iterative methods to compute multiple precision values normally start with low precision then increase precision with each iteration.  For variable length values, this can reduce the cost substantially.  We examine how to do this when computing $\ishinv_h v$.

When computing $\ishinv_h(v)$ from the sequence $\SZ^i(w\iter0)$, only the leading digits of the intermediate iterates matter.    
Rather than compute a series of iterates all of full length, it is possible to compute a sequence of whole inverses, almost doubling their length at each step.   
Theorem~\ref{thm:shift-extension} states this more precisely using $\SZ$ explicitly parameterized by $h$ and $v$, 
\begin{equation}
    \SZ(h, v, w) = w + \Floor{w ( B^h - v w) B^{-h} }.
    \label{eqn:sz-hvw}
\end{equation}
Note that equation \eqref{eqn:sz-hvw} is equivalent to \eqref{eqn:sz} with $u=B^h$.

\begin{Theorem}[Shift Extension]
Let $w = \ishinv_h v$, $B^k \le v < B^{k+1} \le B^h$ and let $w{\ldig n} = \ishift_{n\ell - h+k} (w)$ be the leading $n \ell$ digits of $w$, with $n\ell \le h-k$.
Then
\begin{align*}
 0 \le  w{\ldig 2} - \SZ(k+2\ell, \, v,\, \ishift_\ell w{\ldig 1} ) \le B.
\end{align*}
\label{thm:shift-extension}
\end{Theorem}
\spacetuneclass{acmart}{\vspace{-2\baselineskip}}
\begin{Proof}
Let $\Upsilon =\SZ(k+2\ell, \, v,\,\ishift_\ell w{\ldig 1} )$.
Then, from the definitions and some algebra,
\begin{equation*}
    w{\ldig 2} - \Upsilon =
     \epsilon_1^2  v B^{-k}
     -\epsilon_2
     +\epsilon_3,
\end{equation*}
where 
\begin{equation*}
    \epsilon_1 = \ifrac\!\big (B^{\ell + k}/v\big ), \;
    \epsilon_2 = \ifrac\!\big (B^{2\ell + k}/v\big ), \;
    \epsilon_3 = \ifrac\!\big (B^\ell \epsilon_1  -vB^{-k} \epsilon_1^2\big ).
\end{equation*}
The difference $w{\ldig 2} - \Upsilon$ takes its largest value when $v$ is largest, \textit{i.e.} when $v = B^{k+1}-1$ and
\begin{equation*}
    w{\ldig 2} - \Upsilon \le 
     \epsilon_1^2  B
     -\epsilon_1^2 B^{-k}
     -\epsilon_2 
     +\epsilon_3
     < B+ 1,
\end{equation*}
so $w{\ldig 2} - \Upsilon \le B.$
The difference takes its smallest value when $v$ is smallest, \textit{i.e.} when $v = B^k$ and
\begin{equation*}
    w{\ldig 2} - \Upsilon \ge \epsilon_1^2 -\epsilon_2 +\epsilon_3  \ge -\epsilon_2 > -1,
\end{equation*}
so $w{\ldig 2} - \Upsilon \ge 0$.
\end{Proof}
Compared to the iteration of Theorem~\ref{thm:fast-converge}, this iteration may be one base-$B$ digit short of doubling the precision at each step, rather than being one bit short. This is in trade-off against the savings from working with much smaller values. In any case, iterating \eqref{eqn:sz-hvw}, we have $(\ell-1)2^n + 1$ correct base-$B$ digits after $n$ steps.  It is therefore required to have $\ell \ge 2$ before starting the iteration.

It should be noted that the intermediate expression $B^h - vw$ in \eqref{eqn:sz-hvw} will have about half its digits predictable in advance, since $w$ will be a shifted inverse of $w$. Therefore, in principle, only about half of the digits of the product need be calculated.

\subsection{Divisor Prefixes}
\label{sec:divisor-prefixes}

When the divisor $v$ is large relative to $\ishinv_h v$, its lower order digits will not contribute to the shifted inverse.   
Even if $v$ is not large relative to the final result, it can be large relative to the early short iterates described in Section~\ref{sec:shorter-iterates}.   
It is therefore interesting to see how much a divisor may be perturbed without changing the value of an iterate too much.  
Since Theorem~\ref{thm:shift-extension} shows that short iterates may be one digit short of doubling the precision, we take that as the tolerance here as well.  Using only a prefix of $v$ means dropping some lower order digits, so we are interested in a negative perturbation. This is captured by the following theorem.

\begin{Theorem}[Divisor Sensitivity]
Let $w{\ldig n}$ be as in Theorem~\ref{thm:shift-extension} and let $\Delta$ be the decrease obtained by perturbing the divisor $v$ by $-\delta$ in $\SZ(k+2\ell,\, v,\, \ishift_\ell w{\ldig 1})$, \textit{i.e.}
\begin{equation*}
    \Delta = \SZ(k+2\ell,\, v-\delta,\, \ishift_\ell w{\ldig 1}) - \SZ(k+2\ell,\, v,\, \ishift_\ell w{\ldig 1}).
\end{equation*}
Then \begin{equation}
    B^{2\ell-k-2} \delta - 1 < \Delta < B^{2\ell-k}\delta+1.
    \label{eqn:Delta-inequality}
\end{equation}
In particular,  if $\delta \le B^{k-2\ell + 1}$, then $0 \le \Delta \le B$.
\label{thm:divisor-sensitiviy}
\end{Theorem}
\begin{Proof}
From the definition of $\SZ$ and some simplification, we find
\begin{equation}
    \Delta = \delta {w\ldig 1}^2 B^{-k} +\epsilon_1 - \epsilon_2,
    \label{eqn:Delta-value}
\end{equation}
where
\begin{gather*}
\epsilon_1 = \ifrac\!\big ( w\ldig 1 \big(B^\ell-v w\ldig 1 B^{-k}\big) \big ), \\
\epsilon_2 =\ifrac\!\big ( w\ldig 1 \big(B^\ell-(v\!-\!\delta) w\ldig 1 B^{-k}\big) \big ).
\end{gather*}
Using $B^k \le v < B^{k+1}$ and $w\ldig 1 = \Floor{\floor{B^h/v} B^{\ell-(h-k)}}$, 
equation~\eqref{eqn:Delta-value} 
gives \eqref{eqn:Delta-inequality}
and if $\delta \le B^{k-2\ell+1}$, we have $\Delta < B+1$ so $0 \le \Delta \le B$.
\end{Proof}
Theorem~\ref{thm:divisor-sensitiviy} shows that the last $k-2\ell+1$ digits of $v$ are not required to obtain an iterate with the same order of accuracy as given by a short iterate.
We may therefore adapt the iteration scheme of Theorem~\ref{thm:shift-extension} to be
\begin{align}
    w\iter{i+1} 
    &= \SZ(k+2\ell\iter i-s\iter i,\;
           \ishift_{-s\iter i} v,\;
           \ishift_{\ell\iter i} w\iter i) \\
    &= w\iter i B^\ell + \Floor{
        w\iter i \left ( B^\ell - B^{-k+s\iter i} \floor{v B^{-s\iter i}} w\iter i \right )
    } 
    \nonumber
    \\
    \ell\iter{i+1} &= 2\ell\iter i - 1
    \nonumber
\end{align}
where
\begin{equation}
    s\iter i = \max(0, k-2\ell\iter i + 1).
    \label{eqn:s-equation}
\end{equation}

\subsection{Close Products}
\label{sec:close-products}
When $vw$ is close to $B^h$ the difference $B^h-vw$ will have many fewer than $h$ base-$B$ digits.
When
\begin{equation}
    \big | \,B^h - vw \,\big | \le B^e, \quad e < h,
\end{equation}
only the lower $e$ digits of the product $vw$ need be computed since the upper $h-e$ digits will be determined. 
When $B^h > vw$, the difference will be positive and the upper digits will all be $B-1$.
When $B^h < vw$, the difference will be negative and the upper digits will all be 0.
When the sign of the difference is not known in advance, one may compute the lower $e+1$ digits of the product $vw$ and the sign of the result will be given by whether the coefficient of $B^e$ is 0 or $B-1$.

To compute the lowest $e$ digits of $vw$, one need only compute the lowest $e$ digits
of $(v \irem B^e) \times (n \irem B^e)$.   For some multiplication methods, such as the classical $O(N^2)$ algorithm or the asymptotically faster Karatsuba algorithm, computing the lower digits of this product will be faster than computing the full product by a constant factor.   For other methods, there will be no benefit beyond that provided by having the shorter multiplicands $v \irem B^e$ and $n \irem B^e$.

The value of $e$ will be determined by the precisions $\iprec v = k+1$ and $\iprec w = t+1$, the number $\ell$ of known correct places in $w$  and the required number $g$ of guard digits, as
\begin{equation}
    e \le k + t - \ell + g.
\end{equation}

\section{Integer Algorithm}
\label{sec:integer-algorithm}
Algorithm~\ref{algo:integer-whole-shifted-inverse} presents the results of Section~\ref{sec:base-and-precision-matters} in computational form.   
The initial stages of $\Sc{Shinv}$
(lines \ref{line:shinv-start}-\ref{line:shinv-end-special})
guarantee that the base-$B$ is sufficiently large and that certain easy cases are handled, so $B < v < B^h/2$.   
The next section (lines \ref{line:shinv-start-short}-\ref{line:shinv-end-short}) forms an initial value for the iteration that guarantees fast convergence.  This initial value may have sufficiently many correct digits to be shifted to the required length and returned directly.
otherwise, this initial value is refined in an iteration.

Three variants of refinement are given, incorporating the results of Theorems~\ref{thm:shift-extension} and 
\ref{thm:divisor-sensitiviy} in stages, with $\Sc{Refine3}$ being the method to use in practice.
In each case $w$ is the value that converges to $B^h \iquo v$ and $\ell$ is the number of leading correct base-$B$ digits.
The remainder of the procedure refines the result iteratively using one of the \Sc{Refine} methods.   
Each of these makes use of the \Sc{Step} procedure (lines \ref{line:shinv-start-sz}-\ref{line:shinv-end-sz}), which computes the function given in equation \eqref{eqn:sz-hvw} with the additional parameters $m, \ell$ and $B$.  
Here, $\ell$ is the number of correct leading digits of $w$.  
The parameter $m$ gives the number of additional digits needed, since on the last iteration it will not be necessary to double the number.  This is passed as a parameter so that only the instances of $w$ in \Sc{Step} are shifted, giving smaller products.
All of the \Sc{Refine} methods make use of \Sc{PowDiff} to compute $B^h-v \cdot w$ as described in Section~\ref{sec:close-products} and detailed in Algorithm~\ref{algo:integer-pow-diff}.

\Sc{Refine1} gives a na\"{\i}ve iteration where all computations are performed at the full length of the final result.  By shifting one digit, the iteration avoids
terminating at the $\floor{B^h/v} - 1$ fixed point. No guard digits are needed in the intermediate computation.

\Sc{Refine2} adjusts the iteration so that only the accurate digits of the intermediate results are computed, as described in  Section~\ref{sec:shorter-iterates}.
Two guard digits are required since Theorem~\ref{thm:shift-extension}  shows the short iterate can differ from the truncated full length value by up to $B$.

\Sc{Refine3} additionally uses short divisor prefixes, when possible. 
\Sc{Refine3} is the same as \Sc{Refine2} when $s = 0$.
Two guard digits are required since both the short iterate computation and the use of a divisor prefix can together give a value that is upto $2B$ off from the truncated full length iterate value.
The variable $s$ is as given by equation~\eqref{eqn:s-equation}, taking into account the guard digits.  

Using \Sc{PowDiff} is most beneficial in \Sc{Refine1}, while the use of short iterates and divisor prefixes in \Sc{Refine2} and \Sc{Refine3} already provide a part of this benefit.
A low level implementation would access the base-$B$ digits directly and pre-allocate and re-use a storage region sufficient to hold the largest intermediate results.

The time complexity to compute \Sc{Shinv} depends on the choice of \Sc{Refine} and on the multiplication method used for \Sc{Mult} and \Sc{MultMod}.  
In the following analysis, we assume that the time to compute 
$\Sc{MultMod}(a,b,d,B)$ 
is of the same order as $M(N)$ where $N = \max(\min(\log a,d), \min(\log b,d)))$, \textit{i.e.} that of computing 
$\Sc{Mult}(a \irem B^d,\; b \irem B^d) \irem B^d$.  For FFT multiplication these times will be the same, for other methods they may differ by a constant factor.  

In all cases, \Sc{Shinv} performs $\ceil{\log_2 (h-k)}$ iterations, each having two multiplications.  
When using \Sc{Refine1}, if $k < h/2$,  one multiplication will be of arguments of length between $h-k$ and $h/2$ and the other of length between $h$ and $h-k$,
\textit{i.e.} of time $O(M(h))$ and $O(M(h-k))$.
If $k > h/2$, one multiplication will be of arguments of length $k$ and the other of arguments of length between $h$ and $k$, 
\textit{i.e.} of time $O(M(h))$ and $M(k)$.
Together these give 
$T(h,k) \in O\big (\log(h-k) (M(h) + M(|h/2 -k|))\big )$.  Here we have ignored additive constants.

When using \Sc{Refine2} or \Sc{Refine3} only the necessary prefixes are computed.
With \Sc{Refine3}, at iteration $i$ one multiplication
will be of arguments of length $2^i$.  If $k > h/2$, the second multiplication will be of arguments also of this length.
If $k < h/2$, the second multiplication will be of arguments of 
length $\min(2^i, k)$.  
Again, we have ignored additive constants.
In all cases, letting $h = k + N$,
\begin{equation*}
T(k+N,k) \in O\left ( \sum_{i=1}^{\log N} M(2^i) \right). 
\end{equation*}
This results in time complexity $O(M(N))$ for the 
theoretical 
$M(N) \in O(N \log N)$, for Sch\"{o}nhage-Strassen 
$M(N) \in O(N \log N \log \log N)$ and for practical $M(N) \in O(N^p), p > 0.$

\section{\mbox{Polynomial and Generic Algorithms}}
\label{sec:generic-algorithm}
\subsection*{Polynomial Algorithm}
It is straightforward to adapt Algorithm~\ref{algo:integer-whole-shifted-inverse} to univariate polynomials over a field.
The iteration step function is given by equation \eqref{eqn:polynomial-iteration}.
Algorithm~\ref{algo:polynomial-whole-shifted-inverse} shows the details. 
Polynomial versions of the three \Sc{Refine} methods are shown, but it is \Sc{Refine3} that should be used.

As usual, the polynomial algorithm is simpler than the integer version.
Often iterative methods for polynomials start with a monomial.  Here we start with two terms to simplify the termination condition and the generic algorithm.
Once an initial value is determined, 
the only changes from Algorithm~\ref{algo:integer-whole-shifted-inverse} relate to the consequences of integer arithmetic having carries. 
There is no longer any need for guard digits, nor being one short of doubling precision, nor change of base if $B$ is too small.
Although it is not strictly necessary, we retain the \Sc{PowDiff} operation.  This emphasizes the parallel between the integer and polynomial algorithms, and may provide efficiencies when polynomials are stored densely.

\subsection*{Generic Algorithm}
The benefit of using the whole shifted inverse is that the arithmetic remains in the original domain.  
This allows the iterative algorithm to be defined generically on a domain $D$ with suitable shift, as shown in Algorithm~\ref{algo:generic-whole-shifted-inverse}.  

The generic versions of \Sc{Refine1}, \Sc{Refine2} and \Sc{Refine3}  may be used in place of the function of the same name in Algorithm~\ref{algo:integer-whole-shifted-inverse} or~\ref{algo:polynomial-whole-shifted-inverse}.
It would indeed have been possible to present this generic algorithm first, and show the integer and polynomial cases as specializations, but that would have been less clear.

While the operations \Sc{Mult} and \Sc{MultMod} are mathematically simple, they will typically be implemented by methods provided as procedural parameters.
When there are carries, it is not possible to double the number of correct places at each step and the variable $d$ gives the shortfall. When the arithmetic has  no carries, no guard digits are required.
The complexity analysis of the integer algorithm carries over directly to the polynomial and generic versions.  An implementation should be able to provide $\ishift$ as an $O(1)$ operation.

The domain $D$ need not be commutative. 
It must, however, have a suitable whole shift operation. The shift must be with respect to a central element $b \in D$ (\textit{i.e.}, $bd =db$ for all $d \in D$) and subset $S \subseteq D$ such that every element $d \in D$ can be expressed uniquely as $d = \sum_{i = 0}^k d_i b^i$ with $d_i \in S$. 
In this case, the generic versions of \Sc{Refine1}, \Sc{Refine2} and \Sc{Refine3} all apply.  
An example of such a ring would be the matrix polynomials $D = F^{n\times n}[x]$ with central element $b = x$.
for $u, v$ in such $D$, there exist left and right quotients and remainders $q_L, r_L, q_R, r_R \in D$ such that
\begin{equation*}
    u = v \, q_L + r_L = q_R\, v + r_R
\end{equation*}
with $r = 0$ or $N(r) < N(v)$ for $r \in \{r_L, r_R\}$ and Euclidean norm function $N$ being the degree in $b$.   
The operations $\ishift$ and $\ishinv$ are well-defined and may be used to compute the quotients as
\begin{align*}
 q_L &= \ishift_{-h} (\ishinv_h v \cdot u) &
 q_R &= \ishift_{-h} (u \cdot \ishinv_h v) .
\end{align*}
For polynomials where the variable does not commute with the coefficients, \textit{e.g.} $R[y]\langle \partial_y\rangle$, 
this method applies only in a limited way.
The non-commutative case is discussed further in~\cite{2023-generic-nc-division-preprint}.
\section{Conclusions}
\label{sec:conclusions}

We have shown how to compute whole shifted inverses and quotients for integers and univariate polynomials
with the same order of complexity as multiplication and requiring only domain-preserving ring operations and shifts.   
The algorithms are practical and can be used to modularize software libraries. 
Several results pertaining to the fixed points and convergence of the integer iteration are proven to establish the soundness and efficiency of the algorithm. 
%
We have presented the iterative algorithm generically for domains, not necessarily commutative, endowed with a suitable whole shift operation.

\begin{acks}
We thank Reviewer 3 for a careful reading and colleagues for helpful comments.  This work was supported in part by a grant from the University of Waterloo.
\end{acks}
\bibliography{main}
\newpage
\begin{algorithm}[H]
\caption{$\Sc{Shinv}(v,h,B)$  in $\mathbb Z$\captionstrut}
\label{algo:integer-whole-shifted-inverse}
\begin{algorithmic}[1]
\Require{$ v,  h,B\in \mathbb Z_{>0}$, $B^k \le v < B^{k+1}$}
\Ensure{$\ishinv_{h} v$}
\Comment{All $\ishift$s $\ishinv$s are with respect to $B$}
\Uses{
\Sc{Mult}, a multiplication method\\
\Sc{PowDiff}, to compute $B^h-v w$ (Algorithm~\ref{algo:integer-pow-diff})\\
\Sc{Refine} one of \Sc{Refine1}, \Sc{Refine2} or \Sc{Refine3}}
\Function{\Sc{Shinv}}{$v,h, B$\algomedspace}
\LComment{Group digits if base is small.}\label{line:shinv-start}
  \label{line:shinv-group-start}
\If{$B<16$}
  \State $p \gets \max(6-B,2)$
  \State \Return$\ishift_{h\, \irem \,p-p} \Sc{Shinv}(v,  h \iquo p + 1, B^p)$
  \label{line:shinv-group-end}
\EndIf
\LComment{Special cases  guarantee $B < v \le  B^h/2$.\algosmallspace}
\label{line:shinv-start-special}
\If{$\phantom2 v < B\phantom{^h}$} \Return $B^h \iquo v$ \Comment{Divide by 1 digit} \EndIf
\If{$\phantom2 v > B^h$} \Return $ 0$\EndIf
\If{$2v > B^h$} \Return $1$\EndIf 
\If{$\phantom2 v = B^k$} \Return $B^{h - k}$\EndIf
\label{line:shinv-end-special}

\LComment{Form initial approximation, returning it if sufficient.\algosmallspace}
\label{line:shinv-start-short}
\State $\ell \gets \min(k,2)$
\State $V \gets \sum_{i=0}^\ell v_{k-\ell+i} B^i$
\State $w \gets (B^{2\ell}-V) \iquo V + 1$ \Comment{Divide 4 digits by 2 digits}
\If{$h-k \le \ell$} \Return $\ishift_{h-k-\ell} (w)$ \EndIf
\label{line:shinv-end-short}

\LComment{Refine iteratively using one of the methods below.\algosmallspace}
\State \Return $\Sc{Refine}(v,h,k,w, \ell)$
\EndFunction
\Function{\Sc{Refine1}}{$v, h, k, w, \ell$\algomedspace}
\State $g \gets 1$     
\State $h \gets h + g$
\State $w \gets \ishift_{h-k-\ell}(w)$ \Comment{Scale initial value to full length}
\While{$ h-k >  \ell$}
    \State $w \gets  \Sc{Step}(h, v, w, 0, \ell, 0)$
    \State $\ell \gets   \min(2 \ell - 1,\; h-k)$
    \Comment{Number of correct digits}
\EndWhile
\State \Return $\ishift_{-g}(w)$
\EndFunction
\Function{\Sc{Refine2}}{$v, h, k, w, \ell$\algosmallspace}
\State $g \gets 2$ \Comment{2 guard digits}
\State $w \gets \ishift_g  w$
\While{$ h-k >  \ell$}
    \State $m \gets \min(h-k+1-\ell,\; \ell)$ \Comment{How much to grow}
    \State $w \gets  \ishift_{-1}
      \Sc{Step}\big (k+\ell + m + g,\; v, \;w\;, \ell,\; g \big )$
    \State $\ell \gets \ell + m-1$
\EndWhile
\State \Return $\ishift_{-g}(w)$ 
\EndFunction
\Function{\Sc{Refine3}}{$v, h, k, w, \ell$\algosmallspace}
\State $g \gets 2$ \Comment{2 guard digits}
\State $w \gets \ishift_g  w$
\While{$ h-k >  \ell$}
    \State $m \gets \min(h-k+1-\ell, \,\ell)$
    \State $ s \gets \max(0, \,k - 2\ell + 1 - g)$
    \Comment{How to scale $v$}
    \State $w \gets  \ishift_{-1}
      \Sc{Step}\big (k+\ell + m -s + g,\; \ishift_{-s}\, v,\; w, \; m,\; \ell, \; g \big )$
    \State $\ell \gets \ell + m - 1$  
\EndWhile
\State \Return $\ishift_{-g}(w)$ 
\EndFunction
\Function{\Sc{Step}}{$h, v, w, m, \ell, g$\algosmallspace}
\label{line:shinv-start-sz}
\State $
   \ishift_m w +
   \ishift_{2m-h} \Sc{Mult}
       \big (
          w,
          \Sc{PowDiff}(v, w, h-m, \ell - g, B)
       \big )$\label{line:shinv-end-sz}
\EndFunction
\end{algorithmic}
\end{algorithm}
\newpage
\vbox{
\vspace{-5mm}
\begin{algorithm}[H]
\caption{$\Sc{PowDiff}(v, w, h, \ell, B)$ in $\mathbb Z$\captionstrut}
\label{algo:integer-pow-diff}
\begin{algorithmic}[1]
\Require{$ v, w, h, \ell, B\in \mathbb Z_{>0}$ 
where $\iprec \big |\,w - \ishinv_h v\,\big | \le \iprec w - \ell$
}
\Ensure{$B^h - v\cdot w$}
\Uses{
        $\Sc{Mult}(a,b) = a \cdot b$,\\
        $\Sc{MultMod}(a,b,d,B) = (a\cdot b) \irem B^d$}
\Function{\Sc{PowDiff}}{$v,w,h, \ell, B$\algosmallspace}
\State $L \gets \iprec_B v + \iprec_B w - \ell  + 1$
\If{$v = 0 \vee w = 0 \vee L \ge h$}
   \Return $B^h - \Sc{Mult}(v,w)$
\Else
   \State $P \gets \Sc{MultMod}(v, w, L, B)$
   \If{$P = 0$} \Return 0
   \ElsIf{$P_{L-1} = 0$} \Return $-P$
   \Else ~\Return $B^L - P$
   \EndIf
\EndIf
\EndFunction
\end{algorithmic}
\end{algorithm}
\begin{algorithm}[H]
\caption{$\Sc{Shinv}(v,h)$  in $F[x]$\captionstrut}
\label{algo:polynomial-whole-shifted-inverse}
\begin{algorithmic}[1]
\Require{$v \in F[x], h \in \mathbb Z_{>0}\;$where $k = \iprec v - 1$ and $F$ a field}
\Ensure{$\ishinv_{h} v$} \Comment{All $\ishift$s $\ishinv$s are with respect to $x$}
\Uses{
$\Sc{Mult}(a,b) = a\cdot b$, \\
$\Sc{MultMod}(a,b, d) = (a\cdot b) \irem x^d$, \\
\Sc{Refine} one of \Sc{Refine1}, \Sc{Refine2} or \Sc{Refine3}}
\Function{\Sc{Shinv\,}}{$v,h$\algosmallspace}
\LComment{Special cases.  Afterward $0 < k < h$.}
\If{$k>h$} \Return $ 0$\EndIf
\If{$k=0 \vee k = h \vee v = v_k x^k$} \Return $x^{h - k}/v_k$\EndIf

\LComment{Form initial approximation.\algosmallspace}
\State $w \gets x/v_k - v_{k-1}/v_k^2$;  $\ell \gets 2$

\LComment{Refine iteratively using one of the methods below.\algosmallspace}
\State \Return $\Sc{Refine}(v,h,k,w, \ell)$
\EndFunction
\Function{\Sc{Refine1\,}}{$v, h, k, w, \ell$\algomedspace}
\State $w \gets \ishift_{h-k-\ell}(w)$ \Comment{Scale initial value to full length}
\While{$ h-k + 1 >  \ell$}
    \State $w \gets  \Sc{Step}(h, v, w, 0, \ell)$
    \State $\ell \gets   \min(2 \ell, h-k+1)$
    \Comment{Number of correct digits}
\EndWhile
\State \Return $w$
\EndFunction
\Function{\Sc{Refine2\,}}{$v, h, k, w, \ell$\algosmallspace}
\While{$ h-k + 1 >  \ell$}
    \State $m \gets \min(h-k+1-\ell, \ell)$ \Comment{How much to grow}
    \State $w \gets  
      \Sc{Step}\big (k+\ell + m - 1,\; v, \;w\;, m,\;\ell\big )$
    \State $\ell \gets \ell + m$
\EndWhile
\State \Return $w$
\EndFunction
\Function{\Sc{Refine3\,}}{$v, h, k, w, \ell$\algosmallspace}
\While{$ h-k + 1 >  \ell$}
    \State $m \gets \min(h-k+1-\ell, \,\ell)$
    \State $ s \gets \max(0, \; k - 2\ell + 1)$
    \Comment{How to scale $v$}
    \State $w \gets  
      \Sc{Step}\big (k+\ell + m - 1 - s,\; \ishift_{-s}\, v,\; w, \; m,\; \ell \big )$
    \State $\ell \gets \ell + m$  
\EndWhile
\State \Return $w$
\EndFunction
\Function{\Sc{Step\,}}{$h, v, w, m, \ell$\algomedspace}
\State $
   \ishift_m w +
   \ishift_{2m-h} \Sc{Mult}
       \big (
          w,
          \Sc{PowDiff}(v, w, h-m, \ell)
       \big )$
\EndFunction
\LComment{Compute $x^h - v\cdot w$\algomedspace}
\Function{\Sc{PowDiff\,}}{$v,w,h,\ell$}
\State $L \gets \iprec v + \iprec w - \ell$
\If{$v = 0 \vee w = 0 \vee L \ge h$}
    \Return $x^h - \Sc{Mult}(v,w)$
\Else\
    \Return $-  \Sc{MultMod}(v, w, L)$
\EndIf
\EndFunction
\end{algorithmic}
\end{algorithm}
\vfill
}
\newpage
\vbox{
\vspace{-5mm}
\begin{algorithm}[H]
\caption{Generic \Sc{Refine}s,\; \Sc{Step} and \Sc{PowDiff}\captionstrut}
\label{algo:generic-whole-shifted-inverse}
\begin{align*}
\text{\makebox[0.5\columnwidth][l]{
Certain operations are required on $D$.
On $\mathbb Z$ in base-$B$, these are
}}&\text{\hspace{0.5\columnwidth}}\\
    \ishift_n u &= \floor{u B^n} \\
    \text{coeff}(u, i) &= u_i \\[-0.1\baselineskip]
    \Sc{HasCarries} &= \text{true} \\[-0.1\baselineskip]
    \Sc{Mult}(a, b) &= ab \\[-0.1\baselineskip]
    \Sc{MultMod}(a,b,n) &= ab \irem B^n. 
\\[0.35\baselineskip]
\text{\makebox[0.5\columnwidth][l]{
On $F[x]$, for $F$ a field, these are
}}&\text{\hspace{0.5\columnwidth}}\\[-0.15\baselineskip]
    \ishift_n u &= 
      u \cdot x^n \text{~if~} n \ge 0, \;
      u \iquo x^{-n} \text{~if~} n < 0\\
    \text{coeff}(u, i) &= u[i] \\[-0.1\baselineskip]
    \Sc{HasCarries} &= \text{false} \\[-0.1\baselineskip]
    \Sc{Mult}(a, b) &= ab \\[-0.1\baselineskip]
    \Sc{MultMod}(a,b,n) &= ab \irem x^n.
\end{align*}
\begin{algorithmic}[1]
\LComment{Below, $g$ is no. guard places and $d$ is the prec{.} doubling shortfall.}
\Function{\Sc{D.Refine1\,}}{$v, h, k, w, \ell$\algosmallspace}
\State \algorithmicif\ $\Sc{D.HasCarries}$
    \algorithmicthen\ $g \gets 1; \; d \gets 1$
    \algorithmicelse\  $g \gets 0; \; d \gets 0$
\State $h \gets h + g$
\State $w \gets \mathrm D.\ishift_{h-k-\ell+1-g}(w)$ \Comment{Scale init. value to full length}
\While{$ h-k + 1 - d >  \ell$}
    \State $w \gets  \Sc{D.Step}(h, v, w, 0, \ell)$
    \State $\ell \gets   \min(2 \ell-d, h-k+1)$
    \Comment{Number of correct digits}
\EndWhile
\State \Return $\Sc{D}.\ishift_{-g} w$
\EndFunction
\Function{\Sc{D.Refine2\,}}{$v, h, k, w, \ell$\algomedspace}
\State \algorithmicif\ $\Sc{D.HasCarries}$
    \algorithmicthen\ $g \gets 2; \; d \gets 1$
    \algorithmicelse\  $g \gets 0; \; d \gets 0$
    \State $w \gets \mathrm D.\ishift_g w$
\While{$ h-k + 1 -d>  \ell$}
    \State $m \gets \min(h-k+1-\ell, \ell)$ \Comment{How much to grow}
    \State $w \gets  
      \mathrm D.\ishift_{-d} \, \Sc{D.Step}\big (k+\ell + m +d - 1 + g,\, v, \,w,\, m,\, \ell - g\big )$
    \State $\ell \gets \ell + m - d$
\EndWhile
\State \Return $\Sc{D}.\ishift_{-g} w$
\EndFunction
\Function{\Sc{D.Refine3}\,}{$v, h, k, w, \ell$\algomedspace}
\State \algorithmicif\ $\Sc{D.HasCarries}$
    \algorithmicthen\ $g \gets 2; \; d \gets 1$
    \algorithmicelse\  $g \gets 0; \; d \gets 0$
\State $w \gets \Sc{D}.\ishift_g w$
\While{$h-k+1-d > \ell$}
    \State $m \gets \min(h-k+1-\ell, \,\ell)$
    \State $s \gets \max(0, \; k - 2\ell + 1 - g)$
    \State $t \gets k+\ell + m -s +d - 1 +g$
    \State $w \gets   \Sc{D}.\ishift_{-d} \big (
      \Sc{D.Step}\big ( t
      ,\, \Sc{D}.\ishift_{-s} v,\, w, \, m,\, \ell -g\big )
      \big )$
    \State $\ell \gets \ell + m -d$  
\EndWhile
\State \Return $\Sc{D}.\ishift_{-g} w$
\EndFunction
\Function{\Sc{D.Step}\,}{$h, v, w, m, \ell$\algomedspace}
\State \Return $ \Sc{D}.\ishift_m w \;\;+ $
\State $\quad\quad
   \Sc{D}.\ishift_{2m-h} \Sc{D.Mult}
       \big (
          w,
          \Sc{D.PowDiff}(v, w, h-m, \ell)
       \big )$
\EndFunction
\Function{\Sc{D.PowDiff}\,}{$v,w,h,\ell$\algomedspace}
\State   $c \gets \algorithmicif\ \Sc{D.HasCarries}$ \algorithmicthen\ 1 \algorithmicelse\ 0 
\State $L \gets \Sc{D}.\iprec v + \Sc{D}.\iprec w \;- \ell + c$
\Comment{$c$ for coeff to peek}
\If {$v = 0 \vee w = 0 \vee L \ge h$}
   \State \Return $\Sc{D}.\ishift_h 1 - \Sc{D.Mult}(v,w)$
\Else   
   \State $P \gets \Sc{D.MultMod}(v, w, L)$
   \If {$\Sc{D.HasCarries} \wedge \Sc{D}.\text{coeff}(P, L-1) \ne 0$}
       \State \Return $\Sc{D}.\ishift_L 1 -  P$
   \Else
       ~\Return $-P$
   \EndIf
\EndIf
\EndFunction
\end{algorithmic}
\end{algorithm}
}
\end{document}